\documentclass[aps,twocolumn,showpacs,amsmath,amssymb,prl]{revtex4-1}

\usepackage{graphicx}
\usepackage{comment}
\usepackage[normalem]{ulem}
\usepackage{color}

\newcommand{\be}{\begin{equation}}
\newcommand{\ee}{\end{equation}}
\newcommand{\bfig}{\begin{figure}}
\newcommand{\efig}{\end{figure}}

\makeatletter
\def\blfootnote{\xdef\@thefnmark{}\@footnotetext}
\makeatother

\begin{document}
\title{Temperature-Dependent and Magnetism-Controlled Fermi Surface Changes in Magnetic Weyl Semimetals
}
\author{Nan Zhang$^{1,\S}$, Xianyong Ding$^{2,\S}$, Fangyang Zhan$^2$, Houpu Li$^1$, Hongyu Li$^1$, Kaixin Tang$^1$, Yingcai Qian$^{1,3}$, Senyang Pan$^{1,3}$, Xiaoliang Xiao$^2$, Jinglei Zhang$^3$}
\author{Rui Wang$^{2}$}
\email{rcwang@cqu.edu.cn}
\author{Ziji Xiang$^{1}$}
\email{zijixiang@ustc.edu.cn}
\author{Xianhui Chen$^{1,4}$}
\email{chenxh@ustc.edu.cn}

\affiliation{
$^1$CAS key Laboratory of Strongly-coupled Quantum Matter Physics, Department of Physics, University of Science and Technology of China, Hefei, Anhui 230026, China\\
$^2$Institute for Structure and Function, Department of Physics and Center for Quantum Materials and Devices, Chongqing University, Chongqing 400044, China\\
$^3$High Magnetic Field Laboratory, Chinese Academy of Sciences, Hefei, Anhui 230031, China\\
$^4$Collaborative Innovation Center of Advanced Microstructures, Nanjing University, Nanjing 210093, China
}

\date{\today}

\begin{abstract}
The coupling between band structure and magnetism can lead to intricate Fermi surface modifications. Here we report on the comprehensive study of the Shubnikov-de Haas (SdH) effect in two rare-earth-based magnetic Weyl semimetals, NdAlSi and CeAlSi$_{0.8}$Ge$_{0.2}$. The results show that the temperature evolution of topologically nontrivial Fermi surfaces strongly depends on magnetic configurations. In NdAlSi, the SdH frequencies vary with temperature in both the paramagnetic state and the magnetically ordered state with a chiral spin texture, but become temperature independent in the high-field fully polarized state. In CeAlSi$_{0.8}$Ge$_{0.2}$, SdH frequencies are temperature-dependent only in the ferromagnetic state with magnetic fields applied along the $c$ axis. First-principles calculations suggest that the notable temperature and magnetic-configuration dependence of Fermi surface morphology can be attributed to strong exchange coupling between the conduction electrons and local magnetic moments.
\end{abstract}


\blfootnote{$^\S$These authors contributed equally to this work.}

\maketitle                   
The Fermi surface (FS), an equipotential surface in the momentum space that marks the discontinuity in the distribution of fermions, is only rigorously defined at zero temperature ($T$) \cite{Luttinger}. The thermal broadening of the distribution function at finite $T$ causes a shift of the chemical potential $\mu$ \cite{Ashcroft}, subsequently changing the size of the FS. While such changes correspond to variations in the frequency ($F$) of quantum oscillations (according to the Onsager relation, $F = \frac{\hbar}{2\pi e}A$, where $A$ is the extremal cross-sectional area of FS) in principle, this thermal correction in $F$ is usually too weak to be detected experimentally \cite{footnote1}. Hence, $F$ is routinely treated as $T$-independent in quantum oscillation experiments. Intriguing exceptions do exist. For example, non-parabolic band dispersion gives rise to an additional ``topological" correction of $T$ dependence of $\mu$, which may gives a frequency shift up to $\Delta F/F \sim 1\%$ \cite{CAMcorrection}. Another case stems from the Stoner picture in itinerant ferromagnets, considering a $T$-dependent $F$ induced by the evolution of exchange splitting that continuously modifies the occupation of two spin-polarized subbands \cite{LonzarichGold,ZrZn2}.

Magnetic topological materials have recently become an intense focus of research. They provide a fertile playground for studying the coupling between magnetic orders and electronic band topology, as the two can change concurrently at topological transitions triggered by alternation of system symmetries \cite{GdPtBi,EuB6,EuCd2As2,MnBi2Te4Mao}. Such coupling can also lead to unusual quantum oscillations whose $F$ depends on $T$ \cite{MnBi2Te4Mao,MnBi2Te4Chu,EuMnBi2,PrAlSi}; the underlying mechanisms are not well understood yet, since in most cases the magnetism is local and thus beyond the Stoner picture. In this Letter, we study the Shubnikov-de Haas (SdH) effect (quantum oscillations in electrical resistivity $\rho_{xx}$) in two Weyl semimetals possessing local magnetism, i.e., NdAlSi and CeAlSi$_{0.8}$Ge$_{0.2}$. We show that in NdAlSi the SdH effect exhibits distinct spectra in three phase regimes with different magnetic configurations, namely the high-$T$ paramagnetic (PM) state, the low-$T$ canted up-down-down (u-d-d) ordered state \cite{NdAlSiNeutron} and the field-induced polarized (FIP) state \cite{NdAlSiMagnetic}. Pronounced $T$ dependence of SdH frequencies can be observed in both the PM and the canted u-d-d states, but ceases to manifest itself in the FIP state. In CeAlSi$_{0.8}$Ge$_{0.2}$, $T$-dependent SdH frequencies occur in the ferromagnetic (FM) state below ordering temperature $T_C$ with $H$ applied along the crystalline $c$ axis, yet are absent in the PM state and the FM state with $H$ in the $ab$ plane. Our first-principles calculations ascribe such complex FS evolution to the strong exchange splitting of the Weyl fermion bands caused by the coupling with local 4$f$ electrons.

\begin{figure}[htbp!]
\centering
\includegraphics[width=0.5\textwidth]{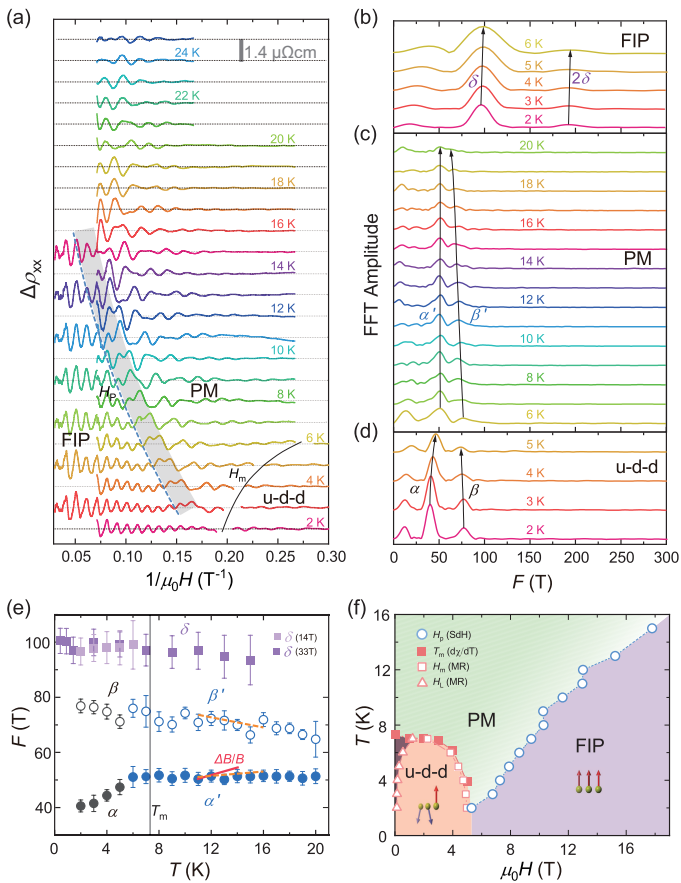}
\caption{(a) The oscillatory resistivity $\Delta \rho_{xx}$ in NdAlSi as a function of inverse field [see Fig.\,S2 in \cite{SM} for raw data]. Data were measured with $H \parallel c$. The thin solid curve marks out the metamagnetic transition field $H_{\rm m}$. Gray thick curve denotes the crossover between the PM and FIP states. (b)-(d) FFT spectra of the SdH oscillations in NdAlSi in the (b) FIP, (c) PM and (d) canted u-d-d states. Arrows guide the eye. (e) Main FFT frequencies plotted against $T$. Error bars are defined as the half FFT peak width at 90$\%$ of the peak height. Dashed and solid lines denote the frequency changes based on the analysis of Lifshitz-Kosevich (LK) fits and shifts of SdH peak positions ($\Delta B/B$), respectively (more discussions are presented in \cite{SM}). (f) $H-T$ phase diagram for NdAlSi obtained from magnetization and MR measurements (Figs.\,S1 and S2 in \cite{SM}). Spin configurations for the u-d-d and FIP states are illustrated. The PM and FIP states are separated here by a threshold field $H_{\rm p}$ for spin polarization [dashed line in (a); see also Fig.\,S2 in \cite{SM}]. The dark shaded area bounded by $T_{\rm m}$ and $H_{\rm L}$ (a low-field jump in MR; see Fig.\,S2 in \cite{SM}) may represent an SDW order \cite{NdAlSiMagnetic}.
}
\label{Fig1}
\end{figure}

Single crystals of NdAlSi and CeAlSi$_{0.8}$Ge$_{0.2}$ were obtained by the flux method (see Sec. I in Supplemental Material \cite{SM}, which includes Refs.\cite{CeAlSiTafti,CeAlSiMOKEPRB,TsuiStark,demagnetizingfactor,Kohn,Kresse1,Kresse2,Perdew,Liechtenstein,Marzari,
Mostofi,Destraz,Shoenberg,Hajdu,SmSb,CeB6,La-CeRu2Si2,ChangRAlGe,LevyAnisotropy,Kavokin}). The two compounds are isostructural: both crystallize in a noncentrosymmetric tetragonal structure [see Fig.\,S1(a) in \cite{SM}] with space group \emph{I}4$_{1}$\emph{md} (No.109), which allows the emergence of Weyl nodes even in the paramagnetic state \cite{NdAlSiNeutron,CeAlSiTafti}. Magnetoresistance (MR) measurements were performed using a standard four-probe configuration in a 14 T superconducting magnet and a 33\,T water-cooled Bitter magnet; as both materials are magnetic, we use $B$ field instead of the applied $H$ field in the analysis of quantum oscillations, considering sample magnetization and the demagnetizing effect (see Sec. II in Supplemental Material \cite{SM}). To analyze the SdH oscillations, fast Fourier transforms (FFT) were performed on the oscillatory MR ($\Delta \rho_{xx}$) that was obtained from a polynomial background subtraction. First-principles and structure calculations were carried out in the framework of density-functional theory (DFT) (see Sec. III in Supplemental Material \cite{SM}).

NdAlSi becomes magnetically ordered at $T_{\rm m}$ = 7.3\,K (Fig.\,S1 \cite{SM}), where it probably enters an incommensurate spin-density-wave (SDW) order before the establishment of a commensurate ferrimagnetic order at 3.3\,K \cite{NdAlSiNeutron,NdAlSiMagnetic}. Both orders manifest a chiral, canted u-d-d spin configuration [see the inset of Fig.\,1(f)]. Because our experimental probe cannot determine the magnetic commensurability, we refer to the low-field magnetic ordering in NdAlSi as the canted u-d-d state in this work. This state terminates at a metamagnetic transition field $H_{\rm m}$ \cite{NdAlSiNeutron,NdAlSiMagnetic} ($\mu_0H_{\rm m} \simeq$ 5.2\,T at 2\,K for $H \parallel c$), which is indicated by a sharp jump in MR (Fig.\,S2, \cite{SM}). At $T$ = 2\,K, the FIP state occurs immediately above $H_{\rm m}$ with Nd 4$f$ moments completely aligned by $H$; at higher $T$, such full polarization is realized at $H_{\rm p} > H_{\rm m}$ (Fig.\,S2, \cite{SM}, note that $H_{\rm p}$ is a characteristic field for a crossover behavior rather than a transition). In Fig.\,1(a) we plot the SdH patterns measured under $H \parallel c$ up to 14\,T at various $T$ (for several temperatures, data up to 33\,T are also shown). Remarkably, single-frequency SdH oscillations [Fig.\,1(b)] appear above $H_{\rm m}$ at $T$ = 2\,K; we take the onset of this feature as the threshold $H_{\rm p}(T)$ for spin polarization. Between $H_{\rm m}(T)$ and $H_{\rm p}(T)$, the SdH patterns resemble that at $T > T_{\rm m}$; thus we assign this field interval to the PM state. An $H-T$ phase diagram is obtained for NdAlSi, as presented in Fig.\,1(f). With increasing $T$, $H_{\rm m}$ and $H_{\rm p}$ decreases and increases, respectively, creating a fan-shaped PM regime in between.

\begin{figure*}[htbp!]
\centering
\includegraphics[width=0.9\textwidth]{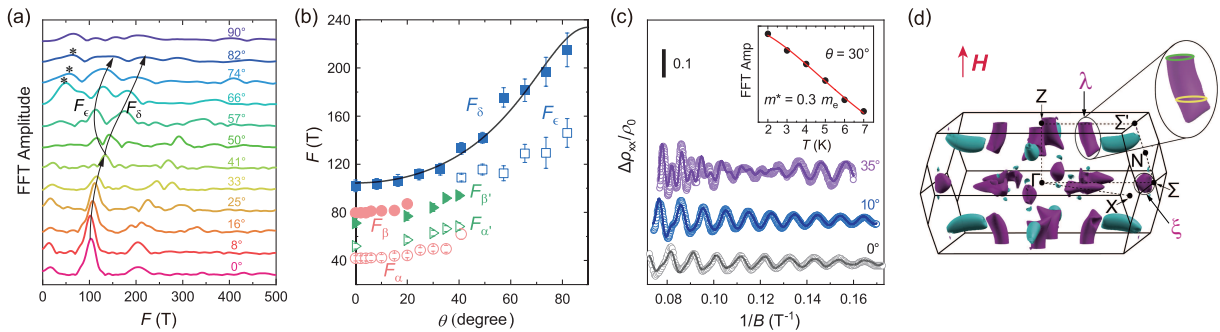}
\caption{(a) FFT spectra of SdH oscillations in NdAlSi measured with varying tilt angle $\theta$ (from $c$ toward $a$ axis) at $T$ = 1.7\,K. FFTs are performed above $H_m$, {\it i.e.,} in the FIP state. (b) SdH frequencies as functions of $\theta$ for the canted u-d-d (circles), PM (triangles), and FIP (squares) states. The solid line is fit to an ellipsoidal FS model (see text). (c) Best fits of the Lifshitz-Kosevich (LK) model (solid lines) to the oscillatory MR in the FIP state (circles) measured at $\theta$ = 0$^\circ$ (black), 10$^\circ$ (blue), 35$^\circ$ (purple), and $T$ = 2\,K. Inset: $T$-dependent SdH oscillation amplitudes measured at $\theta \simeq$ 30$^\circ$ and the LK fit (solid line) (d) DFT-calculated Fermi surfaces (FSs) in the FIP state of NdAlSi. Dark purple and green colors represent hole and electron FS pockets, respectively. An expanded view of the hole pocket ($\lambda$) along the $Z-\Sigma'$ direction is provided. The extremal orbits are highlighted accordingly.
}
\label{Fig2}
\end{figure*}

FFT analysis reveals that the SdH oscillations are composed of two main branches $\alpha$ and $\beta$ in the canted u-d-d state [Fig.\,1(d)], consistent with previous studies \cite{NdAlSiNeutron,NdAlSiMagnetic}. Similarly, two FFT peaks $\alpha'$ and $\beta'$ are resolved in the PM state [Fig.\,1(c)]. In the FIP state, a single component $\delta$ and its second harmonic dominate the SdH pattern [Fig.\,1(b)]. These results corroborate magnetism-controlled FS morphology in NdAlSi. More interestingly, as indicated by the arrows in Figs.\,1(c) and 1(d), the SdH frequencies are $T$-dependent in both the canted u-d-d state and the PM states. The behaviors of $F(T)$ for all SdH branches are summarized in Fig.\,1(e). Branches $\alpha$ and $\beta$ in the canted u-d-d state shift to higher and lower frequencies upon increasing $T$ towards $T_{\rm m}$, respectively: the increase (decrease) of $F_{\alpha}$($F_{\beta}$) from 2\,K ($F_{\alpha}$ = 40\,T, $F_{\beta}$ = 77\,T) to 5 K ($F_{\alpha}$ = 46.5\,T, $F_{\beta}$ = 73.6\,T), corresponds to an FS expansion(shrinkage) of approximately $16 \%$ ($4.5 \%$). Above $T_{\rm m}$, $\alpha$ and $\beta$ smoothly evolve into $\alpha'$ and $\beta'$, suggesting the same origin of the corresponding frequencies; $F_{\alpha'}$ ($F_{\beta'}$) also inherits the $T$-dependence of $F_{\alpha}$ ($F_{\beta}$), though the variations become less remarkable (see Secs. IV and V in Supplemental Material \cite{SM} for detailed analysis). Such $T$-dependent SdH frequencies in the PM state presumably also causes the peculiar oscillations in the $\rho_{xx}(T)$ curves in NdAlSi measured under constant $H$ \cite{NdAlSiRTosc}. In contrast to previous results \cite{NdAlSiNeutron}, we confirm that $F_{\delta}$ $\simeq$ 100\,T in the FIP state does not change with $T$ within our experimental resolution [Fig.\,1(e); see also Figs.\,S2(e) and S2(f) in \cite{SM}].

The fact that the temperature evolution of FSs depends on the magnetic configuration implies an intricate coupling between band structure and magnetism in NdAlSi. To further look into the fermiology, we study the angle-dependent SdH effect. The FFT spectra in the FIP state for varying magnetic-field orientations $\theta$ (angle from $c$ axis toward $a$ axis) are presented in Fig.\,2(a). (Note that $H_{\rm m}$ monotonically increases to $\sim$ 11\,T as $H$ rotates to $\theta \sim$ 70$^\circ$, and is unrecognizable above this angle; see Figs.\,S2(c) and S2(d) in \cite{SM}.) With increasing $\theta$, $F_{\delta}$ becomes higher; for $\theta \gtrsim$ 40$^\circ$, another branch $\epsilon$ appears on the low-frequency side of $\delta$ [arrows in Fig.\,2(a)]. The angle dependence of $F_{\delta}$ can be fitted by an ellipsoidal FS model that is elongated along the $c$-axis with a long-to-short axis ratio of 1.92 [solid line in Fig.\,2(b)]. Moreover, the way the SdH patterns evolves with $\theta$ alludes to changes in spin degeneracy in the corresponding FS. As shown in Fig.\,2(c), with $H \parallel c$, the SdH spectrum in the FIP state can be well described by a single-component Lifshitz-Kosevich (LK) model \cite{Shoenberg,SdH_BEDT-TTF}:
\begin{equation}
\Delta \rho_{xx} = A_{SdH} B^{1/2} \frac{X}{\sinh(X)}\exp(-\frac{\pi m^*}{eB\tau_D})\cos[2\pi(\frac{F_{\delta}}{B}+\phi)],
\label{LK}
\end{equation}
where $X$ = $(2\pi^2k_BTm^*)/e\hbar B$, $m^*$ is the cyclotron mass, $k_B$ the Boltzmann constant, $\tau_D$ the Dingle relaxation time, $\phi$ the phase of SdH oscillations, $A_{\rm SdH}$ an amplitude coefficient, and $F_{\delta}$ = 100\,T. No Zeeman splitting of SdH peaks/valleys is observed at $\theta = 0^\circ$, perhaps implying a spin-polarized FS. When $H$ is tilted from the $c$-axis, signatures indicative of Zeeman splitting appears for $\theta \gtrsim$ 10$^\circ$: the LK fits to SdH patterns at $\theta$ = 10$^\circ$ require the inclusion of the second harmonic for $F_{\delta}$, whereas at $\theta$ = 35 $^\circ$ the fitted amplitude of second harmonic is even larger than the fundamental [Fig.\,2(c); see Supplemental Material \cite{SM} for details]. Such phenomena point toward nearly spin-degenerate bands at higher $\theta$ \cite{footnote2}. Considering the $F-\theta$ relation and the putative spin-polarized nature of branch $\delta$, we assign it to the hole FS pocket along the $Z-\Sigma'$ direction (labeled as “$\lambda$”) in the first Brillouin zone [Fig.\,2(d)]. Due to the damped SdH signals at higher $\theta$ [Figs.\,S2(g) and S2(h) in \cite{SM}], we cannot unambiguously identify all the other branches; discussions of the possible corresponding extremal orbit areas for these frequencies are presented in the Sec. VI of Supplemental Material \cite{SM}. In particular, it is most likely that $\beta$ ($\beta'$) and $\alpha$ ($\alpha'$) also stem from the FS pocket $\lambda$ and correspond to its spin-majority/outer and spin-minority/inner sheets, respectively, in the canted u-d-d (PM) state.

\begin{figure}[htbp!]
\centering
\includegraphics[width=0.45\textwidth]{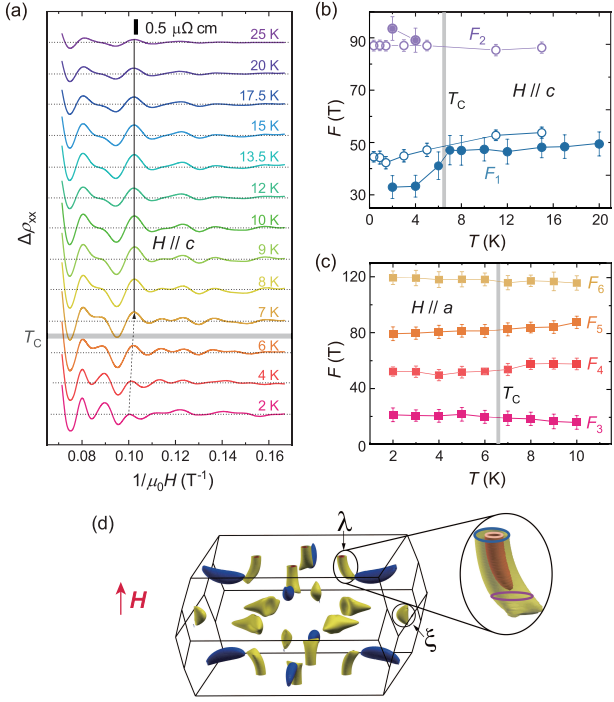}
\caption{(a) $\Delta$$\rho_{xx}$ measured in CeAlSi$_{0.8}$Ge$_{0.2}$ with $H \parallel c$ at different temperatures. Below $T_c$ = 6.5\,K (horizontal line), SdH extrema start to shift with varying $T$ (dotted arrow). (b),(c) $T$ dependence of SdH frequencies (see Fig.\,S3 in \cite{SM} for FFT spectra) for (b) $H \parallel c$ and (c) $H \parallel a$. In (b), the solid and hollow circles are data obtained in two samples $\#$1 and $\#$2 in the field intervals of 8\,T $\leq B \leq$ 14\,T and 8\,T $\leq B \leq$ 33\,T, respectively. Data presented in (c) were measured in sample $\#$1, with FFT performed between 8 and 14\,T. (d) The DFT calculated hole (yellow) and electron (blue) FS pockets for the polarized state with Ce 4$f$ moments aligning along $c$ in an out-of-plane $H$ (arrow). The expanded view shows the outer (spin-majority) and inner (spin-minority) FS sheets for the pocket $\lambda$. Circles highlight the possible extremal orbits.
}
\label{Fig3}
\end{figure}

The $T$-dependent SdH frequencies have been reported in PrAlSi \cite{PrAlSi} but are missing in LaAlSi \cite{LaAlSi}; both are isostructural to NdAlSi and are potential Weyl semimetals. Therefore, the presence of local 4$f$ magnetic moments on the rare-earth site must be the crucial factor inducing such a phenomenon. We verify this by measuring the SdH effect in another isostructural compound with 4$f$ magnetism, CeAlSi$_{0.8}$Ge$_{0.2}$. Among five different members in the series of magnetic Weyl semimetals CeAlSi$_{1-x}$Ge$_x$ (0 $\leq x \leq$ 1), the composition $x$ = 0.2 exhibits the most pronounced SdH effect; see Fig.\,S4 \cite{SM}. This material is ferromagnetic below $T_{\rm C}$ = 6.5\,K; the overall behavior of magnetization (Fig.\,S1 in \cite{SM}) and DFT-determined magnetic structure (Sec.III in Supplemental Material \cite{SM}) are similar to that reported in CeAlSi \cite{CeAlSiTafti}, in which the Ce 4$f$ moments are ordered in an noncollinear ferromagnetic state with in-plane easy axes \cite{CeAlSiMOKEPRB}. Intriguingly, $T$-dependent SdH patterns are only observed with $H \parallel c$ and below $T_{\rm C}$ [Fig. 3(a)], whereas in the PM state above $T_{\rm C}$ and for the FM state in $H \perp c$ they are absent (Fig.\,S3 in \cite{SM}). Figures 3(b) and 3(c) depict the $T$ dependence of FFT frequencies for the SdH measurements under $H \parallel c$ and $H \parallel a$, respectively. $F_1$ measured in the former $H$ orientation is the unique branch that responds remarkably to the variation of $T$: it takes the value of 47\,T (53\,T) above $T_{\rm C}$ in our sample $\#$1($\#$2), yet decreases to 33\,T (45\,T) at 2\,K [Fig.\,3(b)]. All other branches display weak or negligible $T$ dependence [Figs.\,3(b) and 3(c)]. Based on the DFT-calculated extremal orbits, we propose that most of the detected SdH frequencies stem from the FSs $\lambda$ and $\xi$ along the $Z-\Sigma'$ and $\Gamma-\Sigma$ directions, respectively [Fig.\,3(d)]; in particular, the branch $F_1$ is most likely to be associated with a spin-minority pocket (see Sec.VI in Supplemental Material \cite{SM}).

\begin{figure}[htbp!]
\centering
\includegraphics[width=0.48\textwidth]{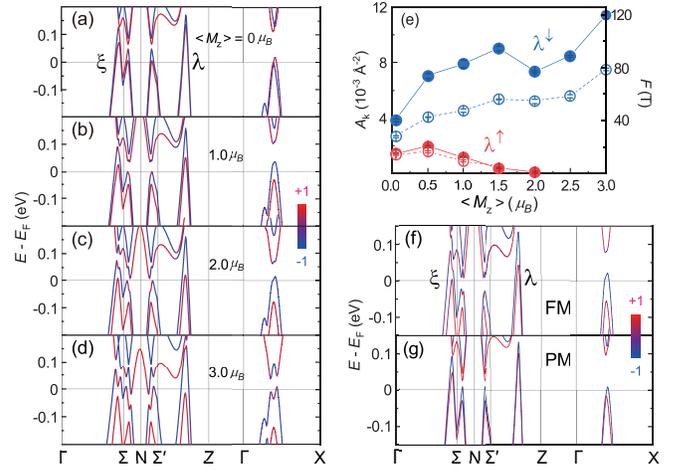}
\caption{(a)-(d) The evolution of spin-resolved band structures of NdAlSi obtained from first-principles calculations; the local moment of Nd$^{3+}$ is constrained to $\langle M_z \rangle$ values of (a) 0, (b) 1.0 $\mu_B$, (c) 2.0 $\mu_B$, and (d) 3.0 $\mu_B$. Note that (a) corresponds to the PM phase under zero field and (d) corresponds to the case of free magnetism (i.e., a FM state from self-consistent calculations). Red and blue colors indicate the $z$ component of spin-up and spin-down states, respectively. (e) The spin-dependent extremal (minimum or maximum) cross-sectional areas on FS pocket $\lambda$, as a function of the polarized local moment, $\langle M_z \rangle$, of the Nd$^{3+}$ ions in NdAlSi. (f) and (g) show the spin-resolved band structures of CeAlSi in the FM (Ce 4$f$ moments align along the $c$ axis) and PM phases, respectively. All calculations include the SOC.
}
\label{Fig4}
\end{figure}

Several mechanism with distinct underlying physics can lead to the temperature-induced FS modification. The topological correction \cite{CAMcorrection} may contribute to but cannot fully account for the large SdH frequency shifts we observed \cite{footnote3}. In Kondo lattices, continuous change of the sizes of FSs with temperature can occur, reflecting the delocalization of $f$ electrons upon cooling due to their hybridization with itinerant $d$ electrons \cite{La-CeRu2Si2}. In the two compounds we study here, however, $f-d$ hybridization is absent and the 4$f$ electrons are completely localized. For instance, in the FIP state in NdAlSi, the cyclotron mass for $F_{\delta}$ is only 0.3 $m_0$ [inset of Fig.\,2(c); $m_0$ is the mass of a free electron], excluding any Kondo-type band renormalization. In Stoner ferromagnets, the exchange splitting of bands scales with the magnetization, thus it is also a function of $T$ \cite{LonzarichGold,ZrZn2}. In NdAlSi and CeAlSi$_{0.8}$Ge$_{0.2}$, this scenario is also inapplicable because the local magnetism herein invalidates the Stoner model. Nonetheless, the $T$ dependence of FSs appears to be sensitive to magnetic configuration, implying that the origin must be the $T$-dependent exchange coupling between the conduction electrons (Weyl fermions) and local 4$f$ moments.

Our DFT calculations show that in both NdAlSi and CeAlSi$_{0.8}$Ge$_{0.2}$ the Weyl fermions at $\epsilon_F$ are predominantly Nd/Ce 5$d$ electrons (Fig.\,S5 in \cite{SM}); they thus have considerable intra-atomic exchange interactions with the local 4$f$ electrons, giving rise to band splitting \cite{ChangRAlGe} that varies with temperature. The total energy splitting of the bands contains three terms: $\Delta E$ = $\Delta_0$ + $E_{ex}$ + $E_z$, where $\Delta_0$ is the zero-field band splitting due to the antisymmetric spin-orbit coupling (SOC) in these noncentrosymmetric materials; $E_{ex}$ and $E_z$ are the exchange splitting~\cite{Hamiltonian} and the Zeeman splitting, respectively. We mention that since both $\Delta_0$ and $E_z$ = $g_s\mu_B B$ ($g_s$ is the Land\'{e} $g$ factor) are independent of $T$, $E_{ex}$ is solely responsible for the observed temperature-induced FS changes \cite{footnote5}. Considering a simplified notion of the exchange splitting: $E_{ex} \propto I_{ex}\langle M_z \rangle$ (where $\langle M_z \rangle$ is the polarized component of the magnetic moment for 4$f^3$ $J$ = 9/2 multiplet along $H \parallel z$), we propose that the $T$-dependent SdH spectrum in the PM state of NdAlSi principally originates from the variation of $\langle M_z \rangle$ at fixed $H$ \cite{CePtBiPRL,CePtBiNJP}.

In real materials, the 4$f$-5$d$ exchange interaction can be much more complicated than the model mentioned above. Nevertheless, our DFT calculations successfully capture the contribution of $E_{ex}$ to the $T$-dependent band structure by tracing its variation upon changing $\langle M_z \rangle$. As displayed in Figs.\,4(a)-4(d), the band splitting in NdAlSi is remarkably enhanced with increasing $\langle M_z \rangle$. In particular, the hole band along the $Z-\Sigma'$ direction [band $\lambda$, Fig.\,2(d)] exhibits nearly two fold degeneracy at $\epsilon_F$ with $\langle M_z \rangle$ = 0 [Fig.\,4(a)]; once the 4$f$ spin polarization is induced by external $H$, the two subbands with opposite $z$-direction spin components split significantly. The spin-minority subband eventually sinks below $\epsilon_F$ for $\langle M_z \rangle > 2 \mu_B$ [Fig.\,4(e)], leaving only one spin-polarized subband that gives the SdH branch $\delta$. This evolution is in agreement with our experimental results. In NdAlSi, we assign the SdH branches $\beta$, $\beta'$ and $\alpha$, $\alpha'$ to the outer and inner FS sheets of band $\lambda$, respectively (Sec. VI in Supplemental Material \cite{SM}); these two groups of $F$ shift toward opposite directions (up and down, respectively) with increasing $\langle M_z \rangle$ upon cooling. DFT calculations qualitatively reproduce such a process [Fig.\,4(e)]. In the FIP state, the inner FS disappears, consistent with our observation of a single branch $F_{\delta}$ which is $T$ independent (due to the saturation of $\langle M_z \rangle$) and is likely to be spin polarized. The fact that $F_{\delta}$ is notably higher than $F_{\beta}$ ($F_{\beta'}$) may reflect a sudden increase of exchange coupling strength upon entering the FIP state \cite{TbSb,NdB6}; a rough estimation based on the DFT-calculated band dispersion yields an enhancement of $E_{ex}$ of $\sim$ 46 meV.

For CeAlSi$_{0.8}$Ge$_{0.2}$, the complex magnetic structure \cite{CeAlSiTafti,CeAlSiMOKEPRB} hinders a direct comparison between theoretical and experimental results. Band structures computed by DFT [Figs.\,4(f) and 4(g)] show that in the fully polarized state with 4$f$ spins aligning along $c$, the band splitting is much larger than that in the PM state. Consequently, it is more reasonable to assign the $T$-dependent SdH branch $F_1$ [Fig.\,3(b)] to an extremal orbit on an inner (minority) FS which shrinks upon increasing spin polarization (Sec. VI in Supplemental Material \cite{SM}). On the other hand, the almost $T$-independent SdH frequencies measured with $H \perp c$ [Fig.\,3(c)] probably imply different responses to the exchange coupling from bands with distinct orbital characters: for $H \parallel c$ ($H \perp c$), the main SdH frequencies arise from band $\lambda$ ($\xi$) (Sec. VI in Supplemental Material \cite{SM}) that is dominated by the $d_{xy}$ and $d_{x^2-y^2}$ ($d_{z^2}$ and $d_{yz}$) orbitals [Fig.\,S5(c) \cite{SM}]. Such complex behavior highlights the influence of SOC in the exchange coupling discussed above, which requires further theoretical investigation to clarify its role. See Sec. VII in Supplemental Material \cite{SM} for more details.

We mention that the exchange-splitting-induced FS changes may explain the T -dependent quantum oscillation frequencies observed in a bunch of magnetic topological materials \cite{MnBi2Te4Mao,MnBi2Te4Chu,SmMnSb2}; though it is usually more significant in the rare-earth compounds \cite{EuCd2As2,EuMnBi2,PrAlSi,EuMnSb2} as a result of large $\langle M_z \rangle$ of the localized 4$f$ electrons. Moreover, it has been pointed out that, with an effective exchange coupling between the localized and itinerant electrons, an indirect Ruderman-Kittel-Kasuya-Yosida (RKKY) interaction between the local moments can be established in topological semimetals, which is mediated by the (partially) spin-polarized Dirac/Weyl fermions \cite{RKKY2015,RKKY2017,RKKY2021}. In a noncentrosymmetric crystal, the antisymmetric SOC can further modify the form of such RKKY interaction, giving rise to chiral spin textures \cite{RKKY2017}; this scenario interprets the origin of the complex magnetic structure in NdAlSi \cite{NdAlSiNeutron}. The experimental evidence for strong localitinerant exchange coupling presented here further verifies the RKKY mechanism and thus helps us understand how the rich magnetic orderings emerge in topological materials.

In summary, we have presented SdH oscillation measurements in different magnetic regimes in Weyl semimetals NdAlSi and CeAlSi$_{0.8}$Ge$_{0.2}$. The SdH frequencies reveal $T$-dependent FS changes that rely on the magnetic configurations: such changes are notable in both the canted u-d-d state and the PM state yet disappear in the high-$H$ FIP state in NdAlSi, whereas they only show up in the FM state with $H \parallel c$ in CeAlSi$_{0.8}$Ge$_{0.2}$. These phenomena can be essentially understood as outcomes of the exchange interactions between the Weyl fermions and the rare-earth 4$f$ local moments, which can persist into the PM state in the presence of finite 4$f$ spin polarization. Our observations of exchange-interaction-induced FS modifications potentially open up a route for realizing manipulation of topological orders in magnetic topological materials.

\vspace{1ex}

We are grateful for the assistance of Chuanying Xi and Yong Zhang in high magnetic field experiments. We acknowledge insightful discussions with Aifeng Wang, Tao Wu, Jianjun Ying and Zhenyu Wang. This work was supported by the National Natural Science Foundation of China (Grants No. 12274390, No. 11888101 and No. 12222402), the Fundamental Research Funds for the Central Universities (WK3510000014), the Strategic Priority Research Program of Chinese Academy of Sciences (XDB25000000), the Innovation Program for Quantum Science and Technology (2021ZD0302802) and Anhui Initiative in Quantum Information Technologies (AHY160000). J.L.Z. was supported by the Excellence Program of Hefei Science Center CAS 2021HSC-UE011. Z.X. acknowledges the USTC startup fund. R. W. acknowledges support by the Beijing National Laboratory for Condensed Matter Physics.


\begin{thebibliography}{99}
\bibitem{footnote1}
For a electronic system with parabolic bands, the Sommerfeld correction of chemical potential $\mu$ at nonzero $T$ can be given by \cite{Ashcroft}
\begin{equation}
\mu (T) = \epsilon_F [1-\frac{1}{3}(\frac{\pi k_BT}{2\epsilon_F})^2],
\label{Sommerfeld}
\end{equation}
thus the relative change of $\mu$ is less than one percent even in a low-carrier-density material with $\epsilon_F \sim$ 25 meV at $T$ = 30\,K. At this temperature, however, the quantum oscillation measurements are mostly hampered by the thermal broadening of Landau levels.
\bibitem{footnote2}
The origin of this phenomenon is not clear yet. It is possible that the large spin splitting in band $\lambda$ in the FIP state (with $H \parallel c$) can be suppressed in a tilted magnetic field. Another explanation is that, with the breaking of the $C_4$ rotational symmetry caused by a tilted $H$, the four $\lambda$ pockets in the first Brillouin zone become inequivalent \cite{CeAlGeSuzuki}. We mention that the shape and size of FSs in the FIP state of NdAlSi may change with field orientation $\theta$ owing to the strong coupling with Nd 4$f$ magnetism. Further theoretical analysis is required to nail down the most likely scenario.
\bibitem{footnote3}
According to the model presented in \cite{CAMcorrection}, the topological correction $\Delta F/F = \Theta(m^*/e\hbar)^2(\pi k_B T/F)^2$, with the maximum $\Theta$ = 5/12. Taking $m^*$ = 0.15 $m_0$ \cite{NdAlSiNeutron}, $F$ = 70\,T for SdH branch $\beta$ in NdAlSi and $T$ = 20\,K, $\Delta F/F$ is at most 0.4$\%$. We cannot verify the presence or absence of such small contibution.
\bibitem{footnote5}
It was pointed out in \cite{NdAlSiNeutron} that the SdH frequency shifts in the canted u-d-d state of NdAlSi are linked to the RKKY interaction between the itinerant Weyl fermions and 4$f$ local moments, with the frequency shifts coinciding with the changes of ordering vectors. Such scenario does not suffice to explain the $T$-dependent SdH patterns  in the PM state.
\end{thebibliography}
\end{document}